\documentclass[12pt,preprint]{aastex}

\usepackage{graphicx}

\shorttitle{Corrective lenses for high-$z$ SNe}
\shortauthors{Dalal, Holz, Chen, \& Frieman}

\newcommand{\figwidth}{0.9\textwidth}

\begin{document}
\title{Corrective lenses for high redshift supernovae}
\author{Neal Dalal}
\affil{Physics Department, University of California at San Diego,
La Jolla, CA 92093}
\author{Daniel E. Holz, Xuelei Chen}
\affil{Institute for Theoretical Physics, University of
California, Santa Barbara, CA 93106}
\and
\author{Joshua A. Frieman}
\affil{NASA/Fermilab Astrophysics Center, Fermilab, PO Box
500, Batavia, IL 60510 and
Department of Astronomy \& Astrophysics, and Center for
Cosmological Physics, University of Chicago, Chicago, IL 60637}

\begin{abstract}
Weak lensing of high-redshift Type Ia supernovae induces an external
dispersion in their observed standard candle brightnesses, comparable
in magnitude to the intrinsic dispersion for
redshifts $z>1$.  The same matter fluctuations responsible for the
magnification of distant supernovae also generate shear in the images
of background galaxies.  We investigate the possibility of using
lensing shear maps constructed from galaxies surrounding the
supernovae as a means of correcting the lensing-induced magnification
dispersion.  We find that a considerable fraction of the lensing
dispersion derives from sub-arcminute scales, which are
not probed by shear maps smoothed on arcminute scales.  We thus find
that weak lensing shear maps will be of only limited value in reducing
the weak lensing magnification fluctuations of supernovae.
\end{abstract}

\section{Introduction}

High redshift Type Ia supernovae provide an excellent means
of studying the expansion history of the
universe~\citep{hzss,scp}. It is estimated that the
intrinsic dispersion in supernova luminosities can be
calibrated to $\approx0.15$ mag, and perhaps in the future
to $0.1$ mag~\citep{snap}, making
them excellent standard candles.  For supernovae at
redshifts $z<1$, this intrinsic dispersion sets the limiting
accuracy with which supernovae may be used to measure
distances.  For higher redshifts ($z\ga1$), however,
gravitational lensing by random fluctuations in the
intervening matter distribution induces a dispersion in
supernova brightness comparable to the
intrinsic dispersion
\citep{frieman97,hw98,holz98,wang}, degrading their value as
standard candles. 
These magnification fluctuations have zero mean, and so may be averaged
away with sufficient numbers of supernovae \citep{wangbin}.  However, the
additional dispersion means that more supernovae are required than
for low redshift samples to achieve a given signal to noise.

It would be of great utility to determine the
gravitational lensing magnification of each
individual supernova. This would allow a correction of the
observed brightnesses of the supernovae, and therefore
improve their use as standard candles. 
Such a correction would be equivalent to obtaining a
larger sample of supernovae, for free. In addition,
measuring the gravitational lensing distribution at high
redshift can be an important probe of the dark
matter~\citep{ms99,seljakholz,ben99,holz01}.  One means of
achieving this would be an inspection of the foreground
galaxies for each supernova.  For example, SN1997ff at
$z=1.7$ has several foreground galaxies in its vicinity,
leading to a magnification possibly as large as 0.4
magnitudes~\citep{ibata,sn1997ff,moertsell}.  If the
magnification factor could be accurately estimated from the
foreground galaxy images, then the supernova brightness
could be corrected to its unlensed value.  The correction
factor depends strongly on uncertain properties of the
galaxies' mass distributions (illustrated by the
controversy over the extent of lensing of SN1997ff),
and would miss possibly important contributions
from dark halos.  Furthermore, since such corrections would
primarily shift highly magnified SNe to lower brightnesses,
while leaving demagnified SNe unaffected, it would bias the
resultant Hubble diagram. It is apparent that direct
identification of individual lenses does not robustly determine
the lensing magnification.  It is also
possible to correlate, in a statistical manner, the
foreground galaxy number density close to the lines of sight to
supernovae with the lensing effects on these
supernovae~\citep{ben01}, but these statistical results do
not help us ``correct'' any given individual supernova.

An alternative method for correcting lensing magnification
is to utilize weak lensing maps constructed from shear
measurements of background galaxies. The same matter
fluctuations responsible for the magnification of supernovae
also lead to shearing of galaxy images.  High redshift SNe
are discovered by repeated exposures of wide fields, which
when co-added provide extremely deep images of the galaxies
surrounding the supernovae.  Such deep, wide field images
are well-suited for measurement of weak lensing shear.  It
is thus natural to hope that mass reconstruction from shear
measurements of the surrounding fields might allow for the
correction of weak lensing magnification, restoring the
supernovae to their intrinsic brightnesses.  A perfect
measurement of the shear field at the redshift of a given
supernova would allow for a perfect reconstruction of the
projected mass surface density (modulo the mass-sheet
degeneracy, which should be unimportant for large enough
fields). From this mass surface density it is possible to
calculate the lensing magnification, and therefore perfectly
account for (and correct) the lensing effects on the
observed brightness. Perfect shear maps are unavailable,
however, and therefore our ability to infer the
magnification is compromised.  In this paper, we investigate
how well weak lensing reconstruction can correct the
brightnesses of distant supernovae.

The basic scheme is as follows. A supernova occurs in a
given field, and its peak apparent magnitude is observed
and calibrated, using some variant of the \citet{phillips}
relation. Then the (co-added) field
containing the supernova is used to estimate the local
shear at the supernova's location by averaging over a smoothing
angle $\theta$. The shear map is then
converted to an effective convergence map using some reconstruction
algorithm such as that of \citet{ks}, and the derived convergence is
used to correct the supernova's standard candle brightness. In the
following section we estimate the variance in convergence
for point sources given knowledge of the smoothed shear map,
$\langle\kappa^2\rangle_\gamma$,
which is a direct measure of the improvement such an
approach can offer. We find that useful corrections require
very large background source galaxy densities, and that this
method is therefore of only marginal utility.

\section{Computing the shear-convergence correlation coefficient}

\begin{figure}
\centerline{
\includegraphics[angle=270,width=\figwidth]{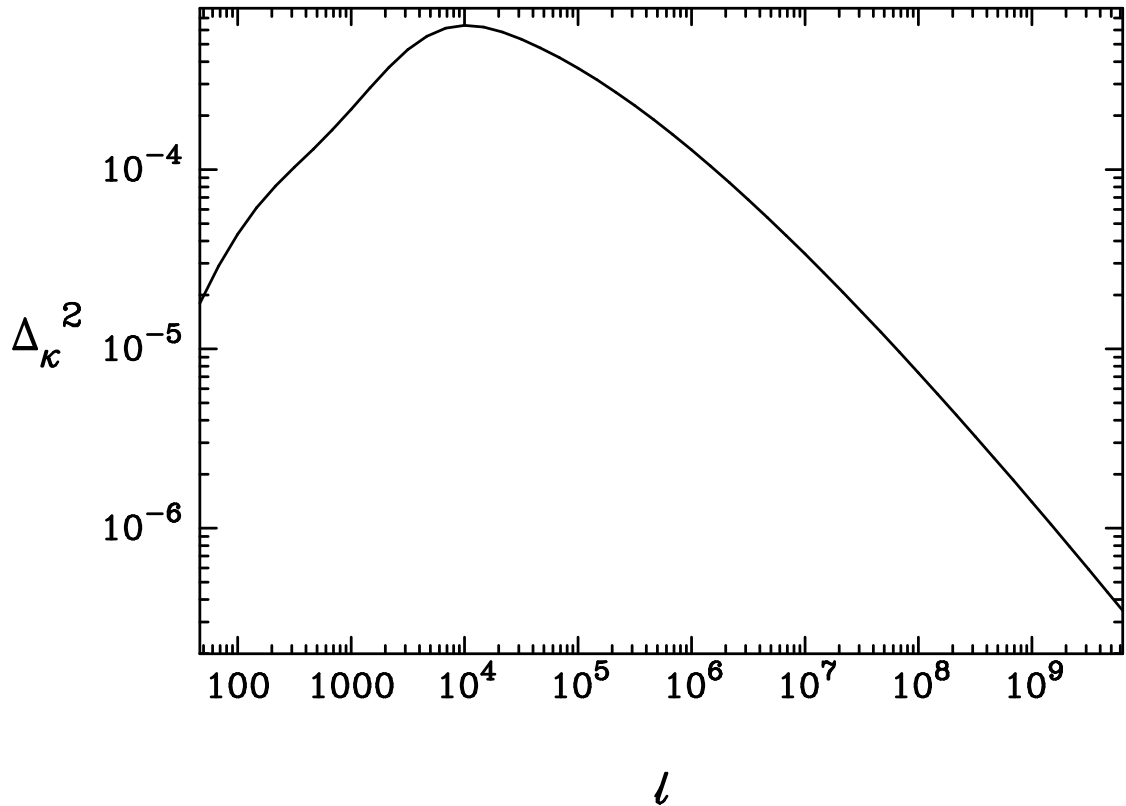}
}
\caption{Weak lensing convergence angular power spectrum at
$z_s=2$, for $\Omega_m=1-\Omega_\Lambda=0.35$, $h=0.65$, $\Omega_b
h^2=0.02$, and $n=1$.  We use the
fitting functions of {\protect\citet{eh99}} for the linear power 
spectrum (COBE normalized), and the fitting functions of {\protect\citet{pd}}
for the nonlinear power spectrum.
Note that the convergence power spectrum peaks at angular
scales of about an arcminute. Smoothing on these scales (or
smaller) will therefore average away much of the convergence power.
}
\label{angpow}
\end{figure}

Let us denote by $\kappa$ the effective convergence,
relative to the homogeneous filled-beam value, for a point
source.  In Figure~\ref{angpow} we plot the angular power
spectrum of the convergence,
$\Delta_\kappa^2(\ell)=\ell^2P_\kappa(\ell)/2\pi$~\citep{whitehu},
for sources at $z_s=2$.  To calculate this, we employ the
fitting functions of~\citet{eh99} for the linear matter
power spectrum, and follow the prescription of~\cite{pd} for
the non-linear correction.  We use a COBE normalized, scale
invariant ($n=1$) linear power spectrum in a flat
$\Lambda$CDM cosmology with total matter density
$\Omega_m=0.35$, Hubble constant $h=0.65$
($H_0=100\,h\,\mbox{km}/\mbox{s/Mpc}$), and baryon density
$\Omega_b h^2=0.02$.  We also assume that the dark matter is
microscopic (e.g., elementary particles), rather than
macroscopic (e.g., black holes or MACHOs). The latter case
leads to enhanced power on microarcsecond scales, which
decorrelates point source magnification from galaxy shear.

The convergence angular power spectrum peaks on arcminute scales
($\ell\sim10^4$), with significant power extending for
multiple decades in $\ell$.  All of this power contributes
to the magnification of (almost) point sources like
supernovae.  When measuring shear, however, galaxy
correlations must be averaged over large angular patches to
suppress Poisson noise, and this averaging
washes out small scale power.  For example, if we smooth
over arcminute-sized patches, we see that a considerable
fraction of the fluctuations affecting the brightness of
point sources are not probed by the smoothed galaxy shear
map.  This hints that shear maps will be of only limited
value.

The convergence power spectrum gives the variance in
effective convergence \citep{whitehu}
\begin{eqnarray}
\left\langle {\kappa^2} \right\rangle&=&{1\over2\pi}\int_0^\infty
\mathrm{d}\ell\,\ell P_\kappa(\ell)\\
&=&\frac{9\pi}{4}\left({\Omega_mH_0^2\over c^2}\right)^2
\int_0^{R_S} \mathrm{d}R\,\left({R(1-R/R_S)\over a(R)}\right)^2
\int_0^\infty {\mathrm{d}k\over k^{2}}\,\Delta_{\rm
mass}^2(k,a(R)),
\label{kappa2}
\end{eqnarray}
where $R$ is
the comoving radial distance
($R(z)=\int_0^z\mathrm{d}z'\,c/H(z')$), $a=1/(1+z)$, and $\Delta_{\rm
mass}^2(k,a)=k^3\,P_{\rm mass}(k)/(2\pi^2)$ is the matter power
spectrum (per logarithmic interval physical
wavenumber). Here, and in what follows, we restrict
ourselves to cosmologies with flat spatial sections ($\Omega_{\rm tot}=1$).
For the power spectrum shown in Figure~\ref{angpow} we find $\left\langle
{\kappa^2} \right\rangle=0.0036$ for sources at $z_s=2$.
In the weak lensing limit, the magnification of a given
source, $\mu$, is related to the convergence by
$\mu\simeq1+2\kappa$. This variance in $\kappa$ thus corresponds
to a $1\sigma$ spread in standard candle flux of $12\%$.

We define $\kappa_\theta$ as the effective convergence
averaged over a circular patch on the sky of radius
$\theta$, which is to be determined by shear
measurements (e.g., using the algorithm of \citet{ks}). 
If $\langle\kappa^2\rangle$ is
the variance in $\kappa$, then by measuring galaxy shear we
can reduce the variance in the corrected convergence of the supernova
image to
$\langle\kappa^2\rangle_\gamma=(1-r^2)\langle\kappa^2\rangle$,
where the correlation coefficient (in the absence of shot
noise) is given by $r^2=\langle\kappa\kappa_\theta\rangle^2/
(\langle\kappa^2\rangle\langle\kappa_\theta^2\rangle)$.
This expression strictly holds for a Gaussian convergence
probability distribution function; although the weak lensing
convergence pdf deviates somewhat from Gaussianity~(see,
e.g., \citet{whm02}), we nonetheless expect the above 
expression for $r^2$ to be a reasonable ballpark estimate of the
correction factor.
It may be interesting to explore
whether non-Gaussian pdf's can lead to superior correction.
Assuming we can write the shot noise term contribution to
$\kappa_\theta^2$ as
$C_{P}(\theta)=\gamma_{\rm rms}^2/N$, with $\gamma_{\rm
rms}$ the intrinsic galaxy ellipticity and $N={\bar
n}\pi\theta^2$ the number of source galaxies inside the smoothing
area ($\bar n$ is the background galaxy number density), we have
\begin{equation}
r^2=\frac{\langle\kappa\kappa_\theta\rangle^2}
{\langle\kappa^2\rangle(\langle\kappa_\theta^2\rangle+C_P(\theta))}.
\label{eqncorr}
\end{equation}
For $N\to\infty$ and $\theta\to0$, we have $r^2\to 1$. This is as
expected: with a perfect lensing shear map, we can fully correct
for lensing magnification.

The correlation functions can all be computed using the nonlinear
matter power spectrum \citep{jainseljak,whitehu,waer},
\begin{equation}
\langle\kappa_\theta^2\rangle=\frac{9\pi}{4}\left({\Omega_mH_0^2\over c^2}\right)^2
\int_0^{R_S} \mathrm{d}R\,\left(\frac{g(R)}{a(R)}\right)^2
\int_0^\infty {\mathrm{d}k\over k^{2}}\,\Delta_{\rm
mass}^2\left({k,a(R)}\right) W_2^2(kR\theta),
\label{kappatheta2}
\end{equation}
and
\begin{equation}
\langle\kappa\kappa_\theta\rangle=\frac{9\pi}{4}\left({\Omega_mH_0^2\over c^2}\right)^2
\int_0^{R_S} \mathrm{d}R\,
{R(1-R/R_S)g(R)\over a(R)^2}
\int_0^\infty {\mathrm{d}k\over k^{2}}\,\Delta_{\rm
mass}^2\left({k,a(R)}\right) W_2(kR\theta),
\label{kappakappatheta}
\end{equation}
where $W_2(x)=2J_1(x)/x$, $J_1(x)$ is the
Bessel function of the first kind, and
\begin{equation}
g(R)=R\int_R^{R_{\rm hor}}\mathrm{d}R_s\,\frac{R_s-R}{R_s}w(R_s),
\end{equation}
with $w(R_s)$ describing the radial distribution of the
sources and $R_{\rm hor}$ the comoving distance to the
horizon.  If all of the source galaxies being
utilized for the shear measurements are at the same
redshift as that of the supernova ($z_s=z_{\rm SN}$), we have
$w(R_s)=\delta(R_s-R_{\rm SN})$. For a given background
galaxy density, this case represents the optimal
shear correction to the convergence. For the more realistic case
of source galaxies distributed in redshift, we adopt a
population distribution described by $w(R_s)\propto
R_s^\alpha\exp(-(R_s/R_\star)^\beta)$~\citep{kaiser92,hu_tomo}, 
and set $\alpha=1$, $\beta=4$, and
$R_\star=c/H_0$ corresponding to a mean redshift of
$\bar z\sim1$. Although this galaxy distribution is
simplistic (e.g., modulo an overall scaling
factor it is independent of survey depth and filters), it is
sufficient to indicate the decorrelation arising from the spread in
galaxy redshifts.  
These two cases circumscribe the range of $r^2$ when tomographic
information, such as photometric redshifts, are employed.

\begin{figure}
\plotone{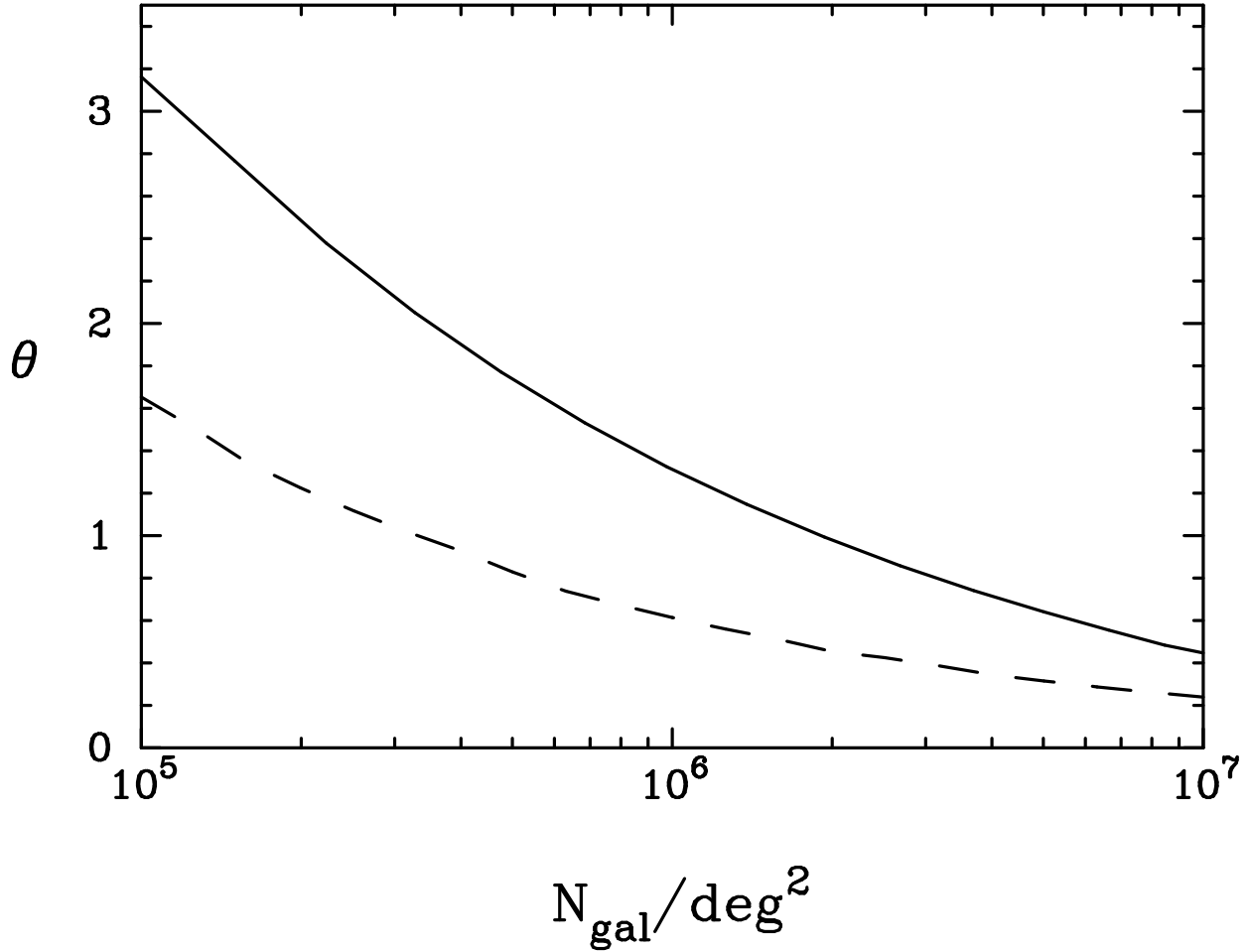}
\caption{The optimum smoothing angle, $\theta$ in arcminutes, as
a function of the background galaxy density, chosen so as to
maximize the correlation coefficient, $r^2$. The supernova
is at redshift $z_{\rm SN}=2$, in a cosmology with
$\Omega_m=1-\Omega_\Lambda=0.35$ and $h=0.65$, and
with an intrinsic galaxy shear of $\gamma_{\rm rms}=0.4$.
The solid line employs the source galaxy distribution
described in the text, while the dashed line represents the
case where all of the source galaxies used to measure shear
are at the same redshift as that of the supernova.}
\label{theta}
\end{figure}

\begin{figure}
\plotone{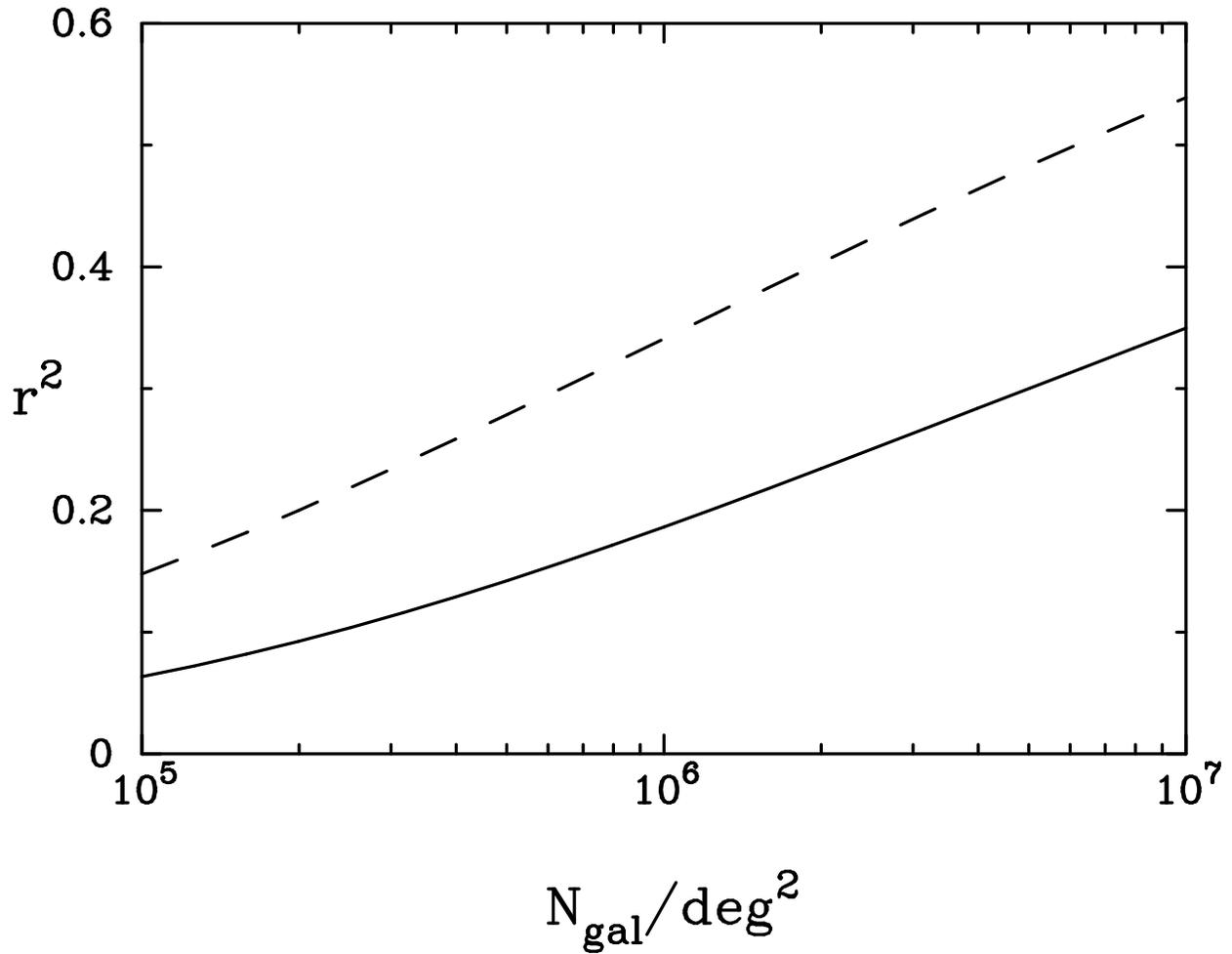}
\caption{The correlation coefficient, $r^2$, given by
equation~({\protect\ref{eqncorr}}), as a function of the
background galaxy density. For each value of galaxy density
the optimal smoothing angle is calculated (shown in
Figure~{\protect \ref{theta}}), and the subsequent shear
correction to magnification is plotted.  As in the previous
Figure, the supernova is at redshift $z_{\rm SN}=2$, in a cosmology
with $\Omega_m=1-\Omega_\Lambda=0.35$ and $h=0.65$,
and with an intrinsic galaxy ellipticity of $\gamma_{\rm
rms}=0.4$.
The solid line employs the source galaxy distribution
described in the text, while the dashed line represents the
case where all of the source galaxies used to measure shear
are at the same redshift as that of the supernova.}
\label{rsq}
\end{figure}

For each value of the background galaxy density there is a
tradeoff between the shot noise term, which decreases for
large smoothing angles, and the cross-correlation term, which is
weaker for large smoothing angles. The smoothing angle,
$\theta$, which maximizes the cross-correlation coefficient,
$r^2$, is shown as a function of $\bar n$ in
Figure~\ref{theta}, for an intrinsic galaxy ellipticity of
$\gamma_{\rm rms}=0.4$~\citep{kaiser92}. The corresponding
values of the cross-correlation coefficient are shown in
Figure~\ref{rsq}. The dashed curves represent the optimal
correction, where all the source galaxies are at the same
redshift as that of the supernova. The solid curves are for
the galaxy redshift distribution described above.

\section{Conclusions}

From Figure~\ref{rsq} it is apparent that shear maps will be
of limited value in reducing the lensing dispersion of
supernova brightnesses, unless the number density of
background galaxies is great enough to permit sufficiently
small smoothing angles.  For example, if the number density
of background galaxies is as high as ${\bar n}=10^6/{\rm
deg}^2$, then at best we find a value for the cross-correlation
coefficient of $r^2\simeq0.35$.  If the uncorrected convergence
variance is
$\langle\kappa^2\rangle=0.0036$, then we find the variance
for the corrected supernovae convergence to be
$\langle\kappa^2\rangle_\gamma=(1-r^2)\langle\kappa^2\rangle=0.0023$. This
yields an rms magnification of $0.1$, which is a 20\%
improvement over the uncorrected value of $0.12$.  This
represents the optimal case discussed above, where all of
the source galaxies are at the same redshift as that of the
supernova. Including the expected spread in galaxy redshifts
provides a more reasonable estimate of $r^2\simeq 0.2$,
giving a reduction in the rms magnification of the supernova
of around $10\%$.  It is to be emphasized that this is an
improvement (reduction) in the width of the observed
supernova magnification distribution, and not a 
change in the mean (which remains at $\mu=1$).  
In addition, the {\em intrinsic}\/
dispersion in supernova luminosities causes a further
contribution to the observed rms standard candle
magnification luminosity.

Tomographic information can do little to ameliorate the
situation.  The simplest approach would be to confine the
shear analysis to source galaxies in a slab in redshift
space centered on the supernova. By doing this one moves up
and to the left of the solid curve in Figure~\ref{rsq},
trading off increased shot noise for more effective lensing
information. It is apparent from the Figure that there is a
net improvement if $\ga10\%$ of the galaxies are at similar
redshifts to that of the supernova.  More
inspired schemes might attempt to employ the information
contained in galaxies at all redshifts; regardless, for a
given effective galaxy density the theoretical limit is
still bounded by the dashed curve in Figure~\ref{rsq}. Our
model for the distribution of galaxy source images in
redshift space is particularly simple---more realistic
models (e.g. with dependence on survey depth) may also push
one closer to the dashed curve.

The estimate presented here is optimistic in that we assume
that the smoothed convergence field may be directly
measured.  In reality, the shear field is measured, and then
converted to a convergence map~\citep{ks}. Even in this
optimistic approximation, at best meager returns are
expected from the construction of shear maps of
surrounding galaxies.
An additional caveat is that we have assumed the noise is
dominated by Poisson noise in the number of source
galaxies. Additional systematic errors, such as imprecise
measurement of the point spread function, only worsen the
decorrelation.
Note that our conclusions are
sensitive to the shape of the convergence angular power
spectrum. If in reality the power spectrum is unlike that of
Figure~\ref{angpow}, and instead has far less small-scale
power, then galaxy shear may turn out to be a much more
powerful tool for correcting weak lensing of supernovae.  At
present, it appears that there is significant small-scale
power \citep{dk1,dk2}, consistent with the values assumed
here.  Future wide field surveys like the
LSST\footnote{http://www.lssto.org} or
SNAP\footnote{http://snap.lbl.gov} will directly measure the
convergence angular power on some of the relevant scales, so
it will be possible to check whether the assumptions made
here are valid.  Assuming, however, that the power spectrum
does not significantly depart from that which we have used,
the prospects for using galaxy shear to correct supernova
brightnesses appear bleak. Given the danger of introducing
unknown biases in the resulting distance-magnitude relation,
it is unclear whether future supernova surveys should
attempt the use of lensing shear maps to correct for
magnification of supernova brightnesses.

\acknowledgments{We thank Eric Linder, Ue-Li Pen, Saul
Perlmutter, Ira Wasserman, and Pengjie Zhang for helpful
discussions.  DEH and XC are supported by the NSF under
grant PHY99-07949 to the ITP. ND is supported by the DOE
under grant DOE-FG03-97-ER 40546, and by the ARCS
Foundation. JF acknowledges support from the DOE,
NASA grant NAG5-10842 at Fermilab, NSF grant PHY-0079251,
and the Center for Cosmological Physics at the University of
Chicago. XC thanks CITA, and ND thanks the ITP, for
hospitality while this work was carried out.}

\end{document}